\documentclass[12pt]{article}
\usepackage{amssymb,amsmath,epsfig}
\usepackage{graphicx}
\usepackage{subfig}
\usepackage{cite}
 \allowdisplaybreaks
\begin{document}
\title{\bf Charged Compact Stars in $f(\mathcal{G})$  Gravity}

\author{M. Ilyas$ $ \thanks{ilyas\_mia@yahoo.com}\\
$ $ Centre for High Energy Physics, University of the Punjab,\\
Quaid-i-Azam Campus, Lahore-54590, Pakistan}

\date{}

\maketitle
\begin{abstract}
This work is devoted to investigate some of the interior configuration of static anisotropic spherical stellar charged structures in the regime of $f(\mathcal{G})$ gravity, where $\mathcal{G}$ is the Gauss Bonnet invariant. The structure of particular charged stars is analyzed with the help of solution obtained by Krori and Barua under different viable models in $f(\mathcal{G})$ gravity theory. The behavior of some physical aspects is investigated with the help of plots and the viability of our modeling is analyzed through different energy conditions. We have also studied some behavior of these realistic charged compact stars and discuss some aspects like density variation, evolution of stresses, different forces, stability of these stars, measure of anisotropy, equation of state parameters and the distribution of charges.
\end{abstract}
\section{Introduction}
Despite of the great, well established and successful theory, the general theory of relativity, in the past century, numerous valuable modifications are being suggested by researchers. In these modification, the Ricci scalar is replace by some arbitrary function, like $f(\mathcal{R})$ in which $\mathcal{R}$ is ricci scalar, $f(\mathcal{G})$, where $\mathcal{G}$ in Gauss-Bonnet invariant and many others as discussed in ref \cite{refa1,refa2,refa3,refa4,refa4a,refa5,refa6,refa7,refa7aaa}.\\
The expansion of the universe is the remarkable phenomenon which is being addressed by these modified theories \cite{ref1}. The well-established fact is, the accelerated expansion of universe cannot be explaining by GR alone in its regular arrangement without adding extra term in the gravitational Lagrangian or exotic matter \cite{ref2,ref3}. The simplest modification was given by Buchdal in 1970 \cite{ref4} with the help of replacement of $\mathcal{R}$ by $f(\mathcal{R})$ arbitrary function of $\mathcal{R}$ in the Hilbert-Einstein gravitational action. In literature, lot of information regarding modified theories of gravity is available \cite{ref5,ref6,ref7,ref8,ref10,ref11,ref12}. One of the well-known modified theory is, Gauss Bonnet gravity, which has been studied many times in the recent past years \cite{ref14,ref15}. In this modified gravity, the Hilbert-Einstein action consist of a function $f(\mathcal{R},\mathcal{G})$ instead of $\mathcal{R}$. This is one of the fact that the additional Gauss-Bonnet term resolve the shortcomings of $f(\mathcal{R})$ gravity theory in the background of large expansion of universe \cite{ref13,ref14,ref15,ref16,ref17}. The simplest form of $f(\mathcal{R},\mathcal{G})$ modified gravity is the $f(\mathcal{G})$ gravity which is widely addressed and can reproduce any kind of cosmological solutions.\\
Like, it could help out in the possible study of an acceleration regimes, and their transition to decelerated regimes, inflationary epoch and passes all tests evoked by solar system experiments and crossing phantom divide line \cite{ref47,ref48}. The $f(\mathcal{G})$ gravity is less constrained than $f(\mathcal{R})$ gravity as discussed in \cite{ref49}. In addition, the $f (\mathcal{G})$ gravity offer an efficient platform to analyze several cosmic issues as an alternate to dark energy \cite{ref50}. similarly, the $f (\mathcal{G})$ gravity is very supportive to study the behavior of finite time future singularities along with late time eras of an accelerating universe \cite{ref52,ref51}. Furthermore, in the background of some viable models in $f(\mathcal{G})$ gravity, the cosmic accelerating nature followed by matter era is also studied \cite{ref49,ref50}. Several viable $f (\mathcal{G})$ gravity models were suggested for the purpose to pass some certain solar system constraints \cite{ref49,ref50} which are studied in \cite{ref53} and further bounds on $f (\mathcal{G})$ gravity models may develop from the behavior of energy conditions \cite{ref54,ref55,ref56}.\\
Observations of compact objects like pulsars, neutron stars and black holes have attracted the researchers towards the useful physical modeling stuff based on highly precise observational data instead of just finding the mathematical expeditions \cite{ref18}.\\
Recently, some of the physical properties of different strange compact stars were studied in the framework of different modified gravity theories and it was concluded that all these strange stars under consideration are stable, matter content is realistic and obeys all the energy conditions \cite{ref19,ref19a,ref19b}.
In favor of modeling static objects, supposition of spherical symmetry geometry is very useful and natural while there are more options in the choice of matter content. In past, many researchers focused their attention on perfect fluid matter content. While fluids with viscosity and pressure anisotropic fluids have also been studied and concluded that the anisotropy disturb the stability of the configuration relative to local isotropic case. Furthermore, the effects of local anisotropy have been elaborated with the help of equation of state \cite{ref21}. Therefore it looks suitable to deal the anisotropic pressure with modified gravity models. Some of the physical properties of compact stars have been studied in the presence of pressure anisotropy and charge \cite{ref25,ref26,ref27,ref28}.\\
The aim of this research work is to investigate the role of $f(\mathcal{G})$ gravity models in modeling of realistic charged compact stellar structures. We investigate the different structural properties, like evolution of charged matter density and anisotropic pressure, the Tolman-Oppenheimer-Volkoff equation, the stability, the equation of state parameters as well as the different energy conditions, for different observational data of compact stars. This paper is design as, in a very next section, we discuss the modified $f(\mathcal{G})$ gravity with charged anisotropic matter distribution of the static spherically symmetric geometry. In section 3, we demonstrate some of viable $f(\mathcal{G})$ gravity models. Section 4 is dedicated to check the physical analysis and viability of different well known compact stars through plots. And finally, we summarize the main results in last section.
\section{$f(\mathcal{G})$ gravity}
This section is to provide the extended version of Gauss-Bonnet gravity with its equations of motion. For $f(\mathcal{G})$ gravity, the usual Einstein-Hilbert action is modified as follows
\begin{equation}\label{action}
S = \int {{d^4}x\sqrt { - g} \left[ {\frac{\mathcal{R}}{2} + f(\mathcal{G})} \right] + {S_m}\left( {{g^{\mu \nu }},\psi } \right) +{S_e}\left( {{g^{\mu \nu }},\psi } \right)},
\end{equation}
where ${\kappa ^2} = 8\pi G \equiv 1$, $\mathcal{R},~f,~{S_m}({{g^{\mu \nu }},\psi}),~{S_e}({{g^{\mu \nu }},\psi})$ are the Ricci
scalar, arbitrary function of Gauss-Bonnet invariant, the matter action and the charged action ,respectively. The Gauss-Bonnet invariant quantity is
\begin{equation}
\mathcal{G} = \mathcal{R}^2 - 4{R_{\mu \nu }}{R^{\mu \nu }} + {R_{\mu \nu \alpha \beta }}{R^{\mu \nu \alpha \beta }},
\end{equation}
where $R_{\mu \nu}$, ${R_{\mu \nu \alpha \beta }}$ are the Ricci and the Riemannian tensors. Upon varying the above action
with respect to $g_{\mu \nu }$, we get the modified field equations for $f(\mathcal{G})$ gravity as
\begin{equation}\label{field equations}
{R_{\mu \nu }} - \frac{1}{2}\mathcal{R}{g_{\mu \nu }} = T_{\mu \nu
}^{\textrm{eff}},
\end{equation}
where $T_{\mu \nu }^{\textrm{eff}}$ is named as effective stress-energy tensor with its expression as follows
\begin{align}\nonumber
T_{\mu \nu }^{{\rm{eff}}} &= {\kappa ^2}({T_{\mu \nu }} + {E_{\mu \nu }}) - 8\left[ {{R_{\mu \rho \nu \sigma }}} \right. + {R_{\rho \nu }}{g_{\sigma \mu }} - {R_{\rho \sigma }}{g_{\mu \nu }} - {R_{\mu \nu }}{g_{\rho \sigma }} + {R_{\mu \sigma }}{g_{\nu \rho }}\\
& + \frac{1}{2}\mathcal{R}({g_{\mu \nu }}{g_{\rho \sigma }} - {g_{\mu \sigma }}\left. {{g_{\nu \rho }})} \right]{\nabla ^\rho }{\nabla ^\sigma }{f_G} + \left( {\mathcal{G}{f_G} - f} \right){g_{\mu \nu }},
\end{align}
where subscript $G$ defines the derivation of the corresponding term
with the GB term, while ${T_{\mu \nu }}$ is the usual stress energy
momentum tensor and
\begin{equation}
 {E_{\mu \nu }} = \frac{{{g_{\mu \mu }}}}{2}\left[ { - {F^{\mu \alpha }}{F_{\alpha \nu }} + \frac{1}{4}\delta _\nu ^\mu {F^{\alpha \beta }}{F_{\alpha \beta }}} \right].
 \end{equation}
\subsection{Anisotropic matter distribution in $f(\mathcal{G})$ gravity}
Here, we wish to examine the effects of anisotropic stresses over the stability of compact charged stars.
For this purpose, we take the distribution of matter content source to be anisotropic having the following mathematical formulation
\begin{align}{T_{\alpha \gamma }} = (\rho  + {P_r}){V_\alpha
}{V_\gamma } - {P_t}{g_{\alpha \gamma }} + \Pi{U_\alpha }{U_\gamma },
\end{align}
where $\rho$ is fluid energy density, $P_t$ is tangential pressure component, $P_r$ is radial pressure component and $\Pi$ is equal to $P_r-P_t$. Furthermore, ${V_\gamma}$ and $U_\gamma$ are four velocity and four vector of the fluid, respectively. These quantities obey ${V^\gamma }{V_\gamma }=1$ and ${U^\alpha }{U_\alpha }=-1$ relation under the comoving coordinate system,.\\
Now, we suppose the interior relativistic structure to be static and spherical symmetric everywhere. In this direction, we take the general line element static spherical symmetric geometry as following
\begin{equation}\label{zz7}
d{s^2} = {e^{a(r)}}d{t^2} - {e^{b(r)}}d{r^2} - {r^2}\left( {d{\theta ^2} + {{\sin }^2}\theta d{\phi ^2}} \right),
\end{equation}
Where $a$ and $b$ are arbitrary constant.
Now by solving the field equations (\ref{field equations}), we get
\begin{align}\nonumber
\rho + E^2  =& \frac{1}{{2{r^2}}}{{\rm{e}}^{ - 2b}}\left[ { - 2{{\rm{e}}^b}} \right. + 2{{\rm{e}}^{2b}} - {{\rm{e}}^{2b}}f{r^2} + {{\rm{e}}^{2b}}{r^2}{f_G}\mathcal{G}\nonumber\\
& + 2b'({{\rm{e}}^b}r - 2({{\rm{e}}^b} - 3){f_G}^\prime ) - 8{f_G}^{\prime \prime } + 8\left. {{{\rm{e}}^b}{f_G}^{\prime \prime }} \right],\label{ro1}\\
{p_r} - E^2 =& \frac{{{{\rm{e}}^{ - 2b}}}}{{2{r^2}}}\left[ {{{\rm{e}}^b}(2} \right. + {{\rm{e}}^b}({r^2}f - 2)) - {{\rm{e}}^{2b}}{r^2}{f_G}\mathcal{G}\nonumber\\
& + 2a'({{\rm{e}}^b}r - 2({{\rm{e}}^b} - 3)\left. {{f_G}^\prime )} \right],\label{pr1}\\
{p_t}+ E^2 =& \frac{1}{{4r}}{{\rm{e}}^{ - 2b}}\left[ { - 2{{\rm{e}}^{2b}}} \right.r{f_G}\mathcal{G} + {{a'}^2}({{\rm{e}}^b}r + 4{f_G}^\prime ) + 2({{\rm{e}}^{2b}}rf - {{\rm{e}}^b}b'\nonumber\\
& + ({{\rm{e}}^b}r + 4{f_G}^\prime ){a^{\prime \prime }}) + a'( - b'({{\rm{e}}^b}r + 12{f_G}^\prime ) + 2({{\rm{e}}^b} + 4\left. {{f_G}^{\prime \prime }))} \right].\label{pt1}
\end{align}
Here, $E^2=\frac{Q^2}{8\pi r^4}$. We suppose $a =r^2 B+C$ and $b=r^2A$ as suggested by Krori and Barua \cite{ref29}, here $A$, $B$ and $C$ are the arbitrary constant. Using these definitions, we reach at
\begin{align}\nonumber
\rho =& 2A{{\rm{e}}^{ - A{r^2}}} - \frac{f}{2} + \frac{1}{{{r^2}}} - \frac{{{{\rm{e}}^{ - A{r^2}}}}}{{{r^2}}} + \frac{1}{2}{f_G}\mathcal{G} - \frac{{{Q^2}}}{{8\pi {r^4}}}+ \frac{{12A{{\rm{e}}^{ - 2A{r^2}}}{f_G}^\prime }}{r}\\
& - \frac{{4A{{\rm{e}}^{ - A{r^2}}}{f_G}^\prime }}{r} - \frac{{4{{\rm{e}}^{ - 2A{r^2}}}{f_G}^{\prime \prime }}}{{{r^2}}} + \frac{{4{{\rm{e}}^{ - A{r^2}}}{f_G}^{\prime \prime }}}{{{r^2}}},\label{ro}\\
{p_r} =& 2B{{\rm{e}}^{ - A{r^2}}} + \frac{f}{2} - \frac{1}{{{r^2}}} + \frac{{{{\rm{e}}^{ - A{r^2}}}}}{{{r^2}}} - \frac{1}{2}{f_G}\mathcal{G} + \frac{{{Q^2}}}{{8\pi {r^4}}}\nonumber\\
&+ \frac{{12B{{\rm{e}}^{ - 2A{r^2}}}{f_G}^\prime }}{r} - \frac{{4B{{\rm{e}}^{ - A{r^2}}}{f_G}^\prime }}{r},\label{pr}\\
p_t=& 2B{{\rm{e}}^{ - A{r^2}}} - A{{\rm{e}}^{ - A{r^2}}} + \frac{f}{2} - AB{{\rm{e}}^{ - A{r^2}}}{r^2} + {B^2}{{\rm{e}}^{ - A{r^2}}}{r^2} - \frac{1}{2}{f_G}\mathcal{G}- \frac{{{Q^2}}}{{8\pi {r^4}}}\nonumber\\
& + \frac{{4B{{\rm{e}}^{ - 2A{r^2}}}{f_G}^\prime }}{r} - 12AB{{\rm{e}}^{ - 2A{r^2}}}r{f_G}^\prime  + 4{B^2}{{\rm{e}}^{ - 2A{r^2}}}r{f_G}^\prime  + 4B{{\rm{e}}^{ - 2A{r^2}}}{f_G}^{\prime \prime }\label{pt}
\end{align}
We will use these equations with different models.
Here, we see charge contribute in $\rho$, $p_r$ and $p_t$. Now consider the quark matter EoS
\begin{equation}
{p_r} = \frac{1}{3}\left[ {\rho  - 4{B_g}} \right]
\end{equation}
where $B_g$ is bag constant. Using this equation, we find the expression for charge, read as
\begin{align}\nonumber
Q=&2{{\rm{e}}^{ - A{r^2}}}\sqrt \pi  r\left[ {{{\rm{e}}^{2A{r^2}}}} \right.(2 - 2Bg{r^2} + {r^2}( - f + \mathcal{G}f')) + 2{f^{\prime \prime }}(3(A - 3B)r\mathcal{G}' - {\mathcal{G}^{\prime \prime }})\nonumber\\
&- 2{{\mathcal{G}'}^2}{f^{\left( 3 \right)}} + {{\rm{e}}^{A{r^2}}}( - 2 + (A - 3B){r^2} + 2{f^{\prime \prime }}( - \left( {A - 3B} \right)r\mathcal{G}' + {\mathcal{G}^{\prime \prime }}) + 2{{\mathcal{G}'}^2}{\left. {{f^{\left( 3 \right)}})} \right]^{\frac{1}{2}}}
\end{align}
The expression for $Q$ contain a square-root which means both sign for charge are acceptable but we will consider the positive sign of charge for further investigation.
\section{Matching condition and Different Models}
In this section, we consider a hypersurface $\Sigma$ that is a boundary of both exterior and interior regions. Furthermore, we suppose Reissner-Nordström metric for the description of exterior geometry, written as
\begin{equation}
d{s^2} = \left[ {1 - \frac{{2m}}{r} + \frac{{{Q^2}}}{{{r^2}}}} \right]d{t^2} - {\left[ {1 - \frac{{2m}}{r} + \frac{{{Q^2}}}{{{r^2}}}} \right]^{ - 1}}d{r^2} - {r^2}\left( {d{\theta ^2} + \sin {\theta ^2}d{\varphi ^2}} \right)
\end{equation}
Where $m$, $r$ and $Q$ is the mass, radius and charge, respectively. The interior of given metric in Eq. (\ref{zz7}) for the charged fluid distribution join smoothly with the above exterior of Reissner-Nordström metric. By matching the these two geometries at $r=R$ and $m(R)=M$, we get
\begin{equation}
A = -\frac{{ 1}}{{{R^2}}}\ln \left[ {1+\frac{{{Q^2}}}{{{R^2}}} - \frac{{2M}}{R}} \right],
\end{equation}
\begin{equation}
B = \left( {\frac{M}{R^3} + \frac{{{Q^2}}}{{{R^4}}}} \right){\left[ {1  + \frac{{{Q^2}}}{{{R^2}}}- \frac{{2M}}{R}} \right]^{ - 1}},
\end{equation}
\begin{equation}
C = \ln \left[ {1  + \frac{{{Q^2}}}{{{R^2}}}- \frac{{2M}}{R}} \right] - \left( {\frac{M}{R} - \frac{{{Q^2}}}{{{R^2}}}} \right){\left[ {1  + \frac{{{Q^2}}}{{{R^2}}}- \frac{{2M}}{R}} \right]^{ - 1}}.
\end{equation}
\begin{table}[h!]
\centering
\begin{tabular}{|c| c| c| c| c| c| c|}
 \hline
Compact Stars  &$M$ & $R(km)$ & $\mu_M=\frac{M}{R}$ &$A$ &$B$ &$\mu_C=\frac{Q^2}{R^2}$ \\ [0.5ex]
\hline\hline\
Vela X - 1 (CS1)  & $1.77M_{\odot}$ & $9.56$ & $0.273091$ & $0.00832706 $ &$0.00608302 $& $0.0133624$\\ [1ex]
\hline\
SAXJ1808.4-3658 (CS2)& $1.435M_{\odot}$ & $7.07$ & $0.299$ & $0.0169456$ &$0.0127081$ & $0.0266898$\\ [1ex]
\hline
4U1820-30 (CS3)& $2.25M_{\odot}$ & $10$ & $0.332$ & $0.00760739 $ &$0.00555676$ & $0.0133208$\\ [1ex]
\hline
\end{tabular}
\caption{The approximate values of the masses, radii and compactness for charged compact stars, Vela X - 1, SAXJ 1808.4-3658, and 4U 1820-30 and their numerical values of the constants $A$, and $B$.}
\label{table:1}
\end{table}
We find the numerical values of these constat for three different strange compact physical stars as shown in Table: \ref{table:1}.
Furthermore, we will take some of the viable models for the study of different compact star properties like the stability analysis energy conditions etc
As $$f(\mathcal{G})=f_i(\mathcal{G})$$
where we will take three different models $i=1, 2, 3$
\subsection{Model 1}
First, we assume the power-law model with the additional logarithmic correction term \cite{ref43}
\begin{equation}\label{model1}
f_1 = \alpha_1 {\mathcal{G}^{n_1}} + \beta_1 \mathcal{G}\log(\mathcal{G}),
\end{equation}
where $\alpha_1,~n_1$ and $\beta_1$ are arbitrary constants. This model could provide observationally well-consistent cosmic results because of its extra degrees of freedom allowed in the dynamics.
\subsection{Model 2}
Next, we take another model having the form \cite{ref44}
\begin{equation}\label{model2}
f_2=\alpha_2  \mathcal{G}^{n_2} \left(\beta_2  \mathcal{G}^m+1\right),
\end{equation}
where $\alpha_2,~\beta_2$ and $m$ are any constant number, while $n_2>0$.
This model is very helpful for the treatment of finite time future singularities.
\subsection{Model 3}
Further, we assume another viable model of the form
\begin{equation}\label{model3}
f_3=\frac{{{a_1} {\mathcal{G}^{n_3}} + {b_1}}}{{{a_2} {\mathcal{G}^{n_3}} + {b_2}}},
\end{equation}
here $a_1$, $a_2$, $b_1$, $b_2$ and $n_3$ are arbitrary constants with $n_3>0$.\\
From the condition $p_r(R)=0$, We find the value of $B_g$ as shown in Table: \ref{table:2}.
\begin{table}[h!]
\centering
\begin{tabular}{|c| c| c| c|}
 \hline
Models  & $B_g$ for CS1 & $B_g$ for CS2 & $B_g$ for CS3 \\ [0.5ex]
\hline\hline\
Model 1 & 0.00336605 & 0.00635613 & 0.00307594\\ [1ex]
\hline\
Model 2 & 0.00336605 & 0.006356117 & 0.003075942\\ [1ex]
\hline
Model 3 & 0.00336603 & 0.006355635 & 0.003075926\\ [1ex]
\hline
\end{tabular}
\caption{The approximate values of the the constant $B_g$ for the three different stars under three different models. }
\label{table:2}
\end{table}
\\
Using these model with eq. (\ref{ro}-\ref{pt}), we get $\rho$, $p_r$ and $p_t$ from which we check the different aspect of compact stars as shown in Table \ref{table:1}. We will discuss these aspect one by one in the following section.
\section{Aspects of $f(\mathcal{G})$ Gravity Models}
In this section, we discuss some of physical aspects of the above charged stars from the interior solution. We present the anisotropic behavior and stability of these charged stars under consideration of three different $f(\mathcal{G)}$ viable models. We discuss these aspects one by one in following
\subsection{Variation of Energy Density and anisotropic stresses}
\begin{figure} \centering
\epsfig{file=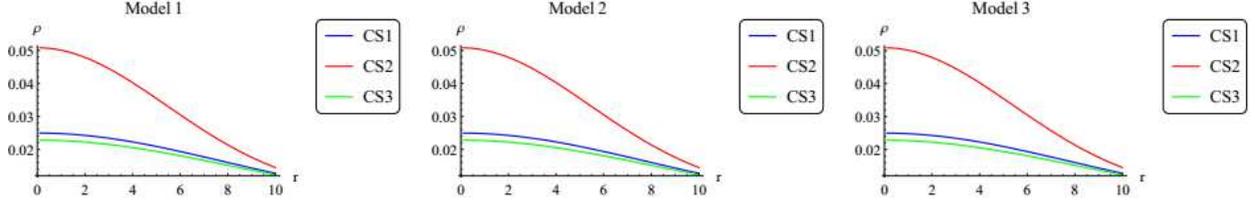,width=1\linewidth}
\caption{Variation of density profile for charged stars, Vela X - 1, SAX J 1808.4-3658, and 4U 1820-30, under different viable $f(\mathcal{G})$ models.}\label{roc}
\end{figure}

\begin{figure} \centering
\epsfig{file=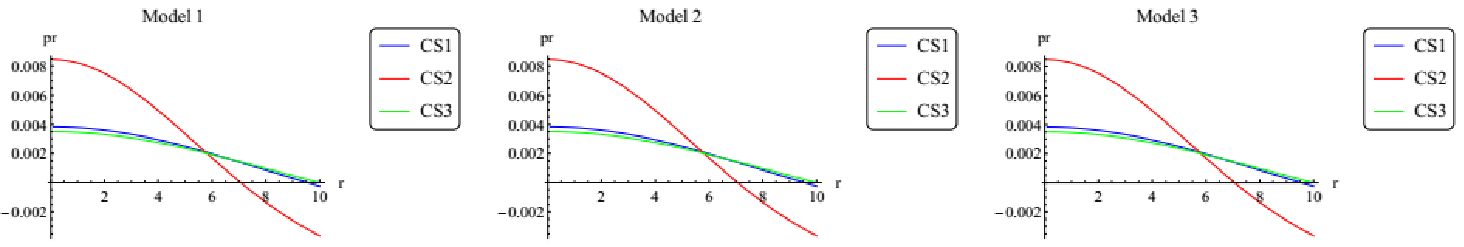,width=1\linewidth}
\caption{Evolution of radial pressure for charged stars, Vela X - 1, SAX J 1808.4-3658, and 4U 1820-30, under different viable $f(\mathcal{G})$ models.}\label{prc}
\end{figure}

\begin{figure} \centering
\epsfig{file=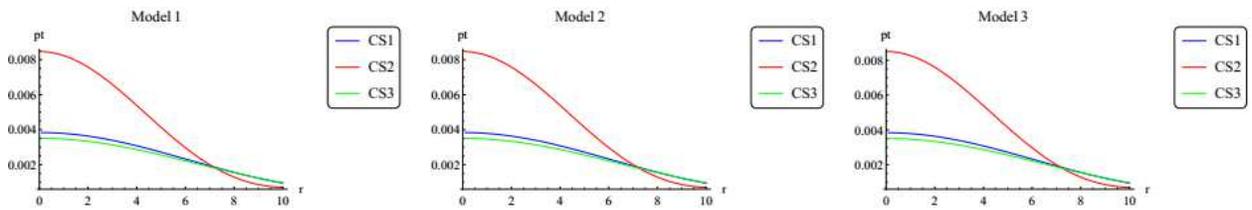,width=1\linewidth}
\caption{Evolution of transverse pressure for charged stars, Vela X - 1, SAX J 1808.4-3658, and 4U 1820-30, under different viable $f(\mathcal{G})$ models.}\label{ptc}
\end{figure}

\begin{figure} \centering
\epsfig{file=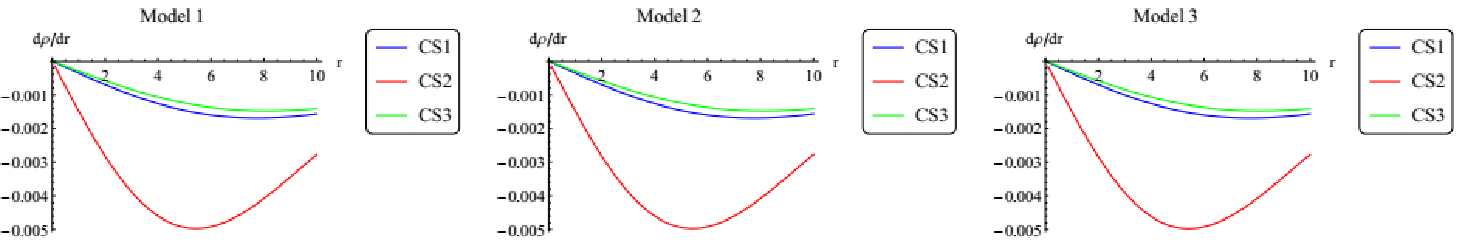,width=1\linewidth}
\caption{Evolution of $d\rho/dr$ for charged stars, Vela X - 1, SAX J 1808.4-3658, and 4U 1820-30, under different viable $f(\mathcal{G})$ models.}\label{dro}
\end{figure}

\begin{figure} \centering
\epsfig{file=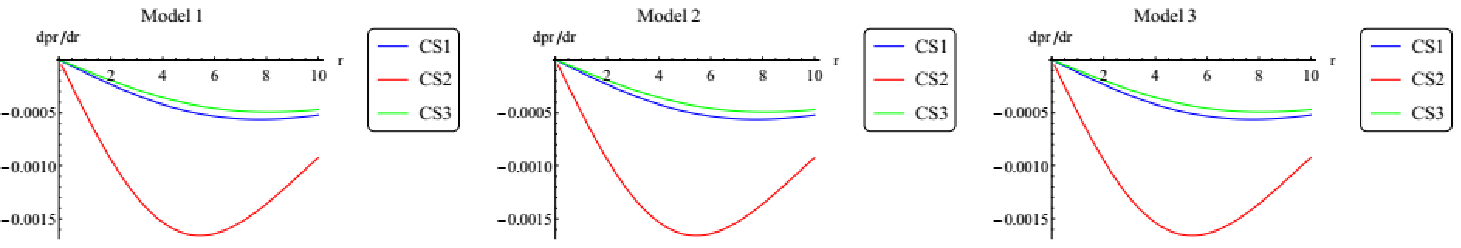,width=1\linewidth}
\caption{Evolution of $dp_r/dr$ for charged stars, Vela X - 1, SAX J 1808.4-3658, and 4U 1820-30, under different viable $f(\mathcal{G})$ models.}\label{dpr}
\end{figure}

\begin{figure} \centering
\epsfig{file=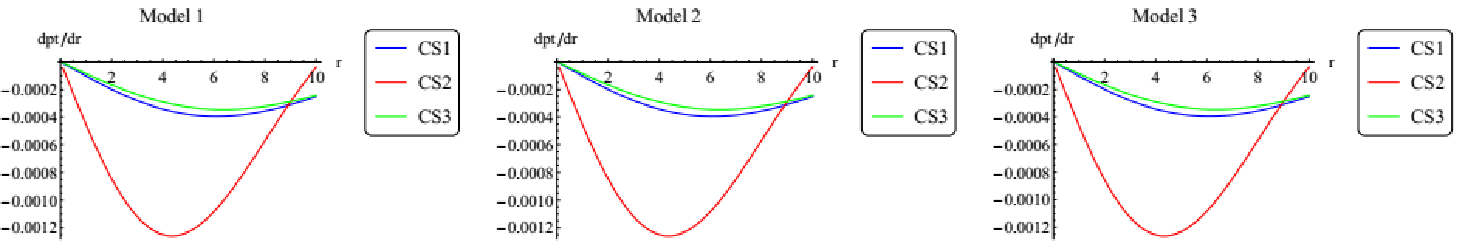,width=1\linewidth}
\caption{Evolution of $dp_t/dr$ for charged stars, Vela X - 1, SAX J 1808.4-3658, and 4U 1820-30, under different viable $f(\mathcal{G})$ models.}\label{dpt}
\end{figure}

We study the influence of quark matter EoS with the anisotropic stresses at the center with modified $f(\mathcal{G})$ gravity models. The corresponding variations in the vicinity of energy density along with anisotropic stresses are shown in Figs. \ref{roc}, \ref{prc} and \ref{ptc}, respectively.\\
The evolution of the density for the strange star candidate $Vela X - 1$, $SAX J 1808.43658$, and $4U 1820-30$ are shown in Fig. \ref{roc}. Here, for $r\rightarrow0$, the density goes to its maximum value. In fact, this indicates the high compactness of the core of these stars and validating our models in $f(\mathcal{G})$ gravity under investigation for
the outer region of the core.\\
In case of model 1: The density at core of Vela X - 1 is $1.33881\times10^{15} gcm^{-3} $, the density at core of SAX J 1808.43658 is $2.72436\times10^{15} gcm^{-3} $, while 4U 1820-30 having density at core is $1.223106\times10^{15} gcm^{-3}$.\\
Furthermore, the surface density of Vela X - 1 under model 1 is $7.215861\times10^{14} gcm^{-3}$, SAX J 1808.43658 having $1.362573\times10^{15} gcm^{-3}$, while 4U 1820-30 having $6.593945\times10^{14} gcm^{-3}$.\\
The Central and surface density of these stars for different models are shown in Table \ref{table:3}.\\
Similarly, the variation of the radial and traverser pressure,are shown in Fig. \ref{prc}-\ref{ptc}.\\
The behavior of radial pressure is, for $r\rightarrow0$, the radial pressure $1.849116\times10^{35} g cm^{-1}sec^{-2}$ for Vela X - 1 under model 1. SAX J 1808.43658 having $4.0796995\times10^{35} g cm^{-1}sec^{-2}$ at core and in case of 4U 1820-30, the radial pressure at core is $1.688794\times10^{35} g cm^{-1}sec^{-2}$.\\
The transverse pressure in consideration of model 1,
For Vela X - 1, the transverse pressure is $1.84911386\times10^{35} g cm^{-1}sec^{-2}$, for SAX J 1808.43658 is $4.0795867\times10^{35} g cm^{-1}sec^{-2}$ while for 4U 1820-30 is $1.6887926204\times10^{35} g cm^{-1}sec^{-2}$.\\
The variation of radial derivative of density, $\frac{d\rho}{{dr}}$, radial derivative of radial pressure, $\frac{dp_r}{{dr}}$, and radial derivative of transverse pressure, $\frac{dp_t}{{dr}}$ are shown in Fig. \ref{dro}, \ref{dpr} and \ref{dpt} respectively.
We see that these variations are negative and for $r=0$, we get
$${\left. {\frac{{d\rho }}{{dr}}} \right|_{r = 0}} = 0$$
$${\left. {\frac{{d{p_r}}}{{dr}}} \right|_{r = 0}} = 0$$
which is expected. e.g. central density of stars $\rho(r=0)=\rho_c$.
\begin{table}[h!]
\centering
\begin{tabular}{|c| c| c| c|}
 \hline
Density of Stars & Model 1& Model 2 & Model 3 \\
\hline\hline\hline\
Vela X - 1 $\rho_c (g/cm^3)$ &$1.33881\times10^{15}$& $1.33881\times10^{15}$& $1.33904\times10^{15}$\\
\hline\
Vela X - 1 $\rho_R (g/cm^3)$ &$7.21586\times10^{14}$ &$7.21586\times10^{14}$& $7.21581\times10^{14}$\\
\hline\hline\
SAX J 1808.43658 $\rho_c (g/cm^3)$ &$2.72436\times10^{15}$ &$2.72449\times10^{15}$& $2.72857\times10^{15}$\\
\hline\
SAX J 1808.43658 $\rho_R (g/cm^3)$ &$1.36257\times10^{15}$ &$1.36257\times10^{15}$ &$1.36247\times10^{15}$\\
\hline\hline\
4U 1820-30 $\rho_c (g/cm^3)$ &$1.22311\times10^{15}$ &$1.22311\times10^{15}$& $1.22327\times10^{15}$\\
\hline\
4U 1820-30 $\rho_R (g/cm^3)$ &$6.59394\times10^{14}$& $6.59394\times10^{14}$& $6.59391\times10^{14}$\\
\hline
\end{tabular}
\caption{The approximate values of Central density $\rho_c$ and Surface density $\rho_R$ for three different models. }
\label{table:3}
\end{table}
\subsection{Energy conditions}
\begin{figure}
\centering
\epsfig{file=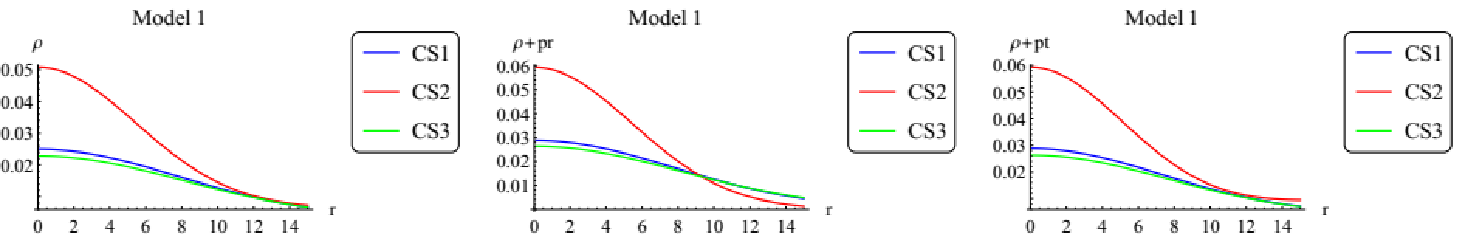,width=1\linewidth}
\epsfig{file=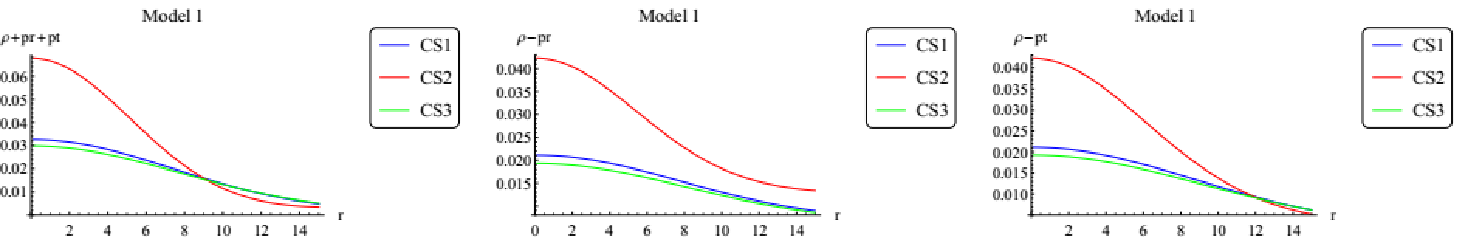,width=1\linewidth}
\caption{Different Energy conditions for Model 1}\label{energy1}
\end{figure}

\begin{figure}
\centering
\epsfig{file=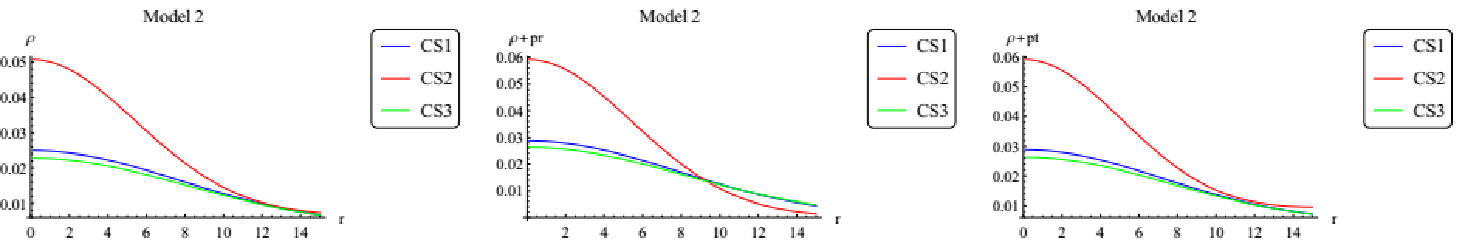,width=1\linewidth}
\epsfig{file=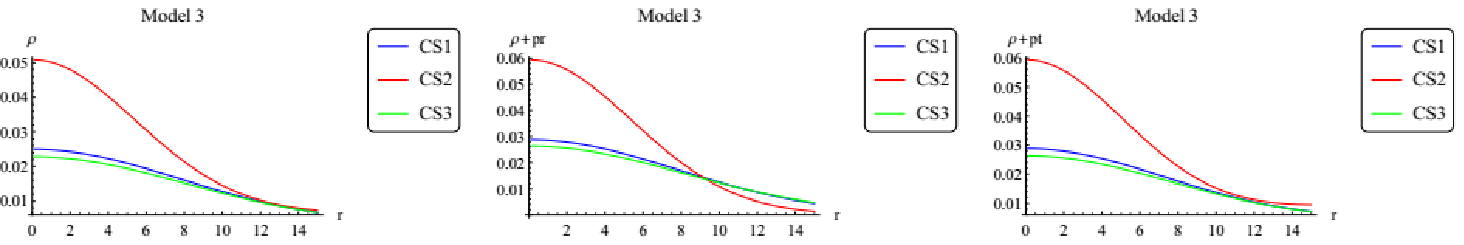,width=1\linewidth}
\caption{Different Energy conditions for Model 2}\label{energy2}
\end{figure}

\begin{figure}
\centering
\epsfig{file=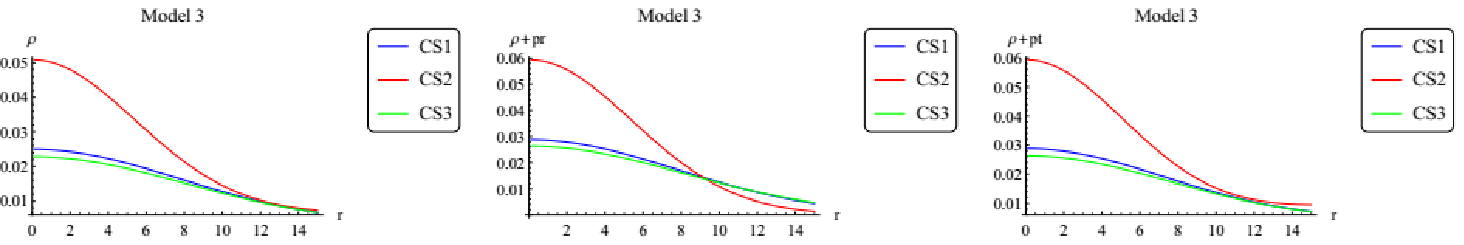,width=1\linewidth}
\epsfig{file=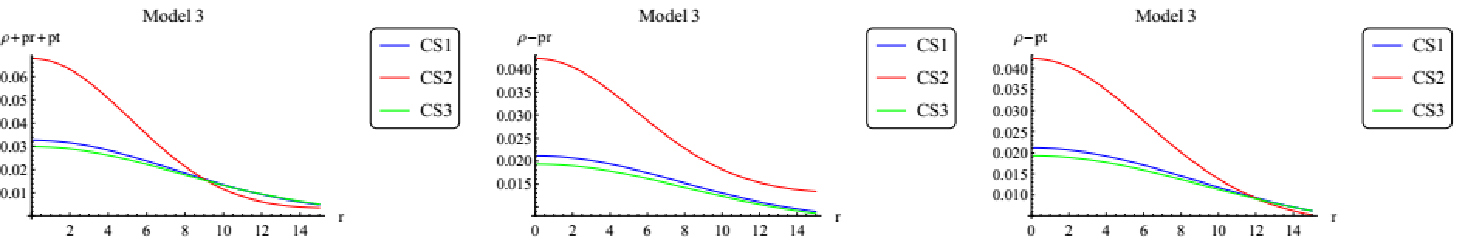,width=1\linewidth}
\caption{Different Energy conditions for Model 3}\label{energy3}
\end{figure}

To deal with a physically viable and acceptable matter field, there are some mathematical constraints which should be obeyed by stress-energy tensor, these constraints are known as energy conditions. These energy conditions are coordinate invariant and can be written as following.\\
\begin{itemize}
\item NEC: $ \rho  + {p_i} \ge 0$ .\\
\item WEC: $ \rho  \ge 0$, $ \rho  + {p_i} \ge 0$ .\\
\item SEC: $ \rho  + {p_i} \ge 0$, $ \rho  + {p_i} + {p_t} \ge 0$ .\\
\item DEC: $ \rho  \ge |{p_i}|$ .
\end{itemize}
Here $i=r,t$ and $\rho$, $p_r$ and $p_t$ include electric charge contributions as well.\\
All these above energy conditions for three different charged compact relativistic
structures are well satisfied under consideration of different viable $f(\mathcal{G})$ gravity models.
The evolution of these energy conditions are shown graphically in Fig. \ref{energy1}, \ref{energy2} and \ref{energy3}.\\
\subsection{Equilibrium condition}
To investigate the equilibrium of inner structure of these charged compact stars, we use the generalized Tolman-Oppenheimer-Volko (TOV) equation. For charged spherical anisotropic stellar interior geometry, this equation is written
\begin{equation}\label{tov1}
\frac{{d{p_r}}}{{dr}} + \frac{{\nu'(\rho  + {p_r})}}{2} + \frac{{2({p_r} - {p_t})}}{r} + \frac{{\sigma Q}}{{{r^2}}}{e^{\lambda /2}} = 0
\end{equation}
Where $\sigma$ is charge density. Furthermore, the above Eq. (\ref{tov1}) may be written as a sum of different forces e.g. gravitational, hydrostatic, anisotropic and electric forces
\begin{equation}
F_g + F_h + F_a+F_e= 0,
\end{equation}
which yields
$$F_g=- r B(\rho+p_r),  F_h=\frac{{-d{p_r}}}{{dr}},   F_a= 2\frac{{({p_r} - {p_t})}}{r}, F_e= \frac{{\sigma Q}}{{{r^2}}}{e^{\lambda /2}}$$
\begin{figure} \centering
\epsfig{file=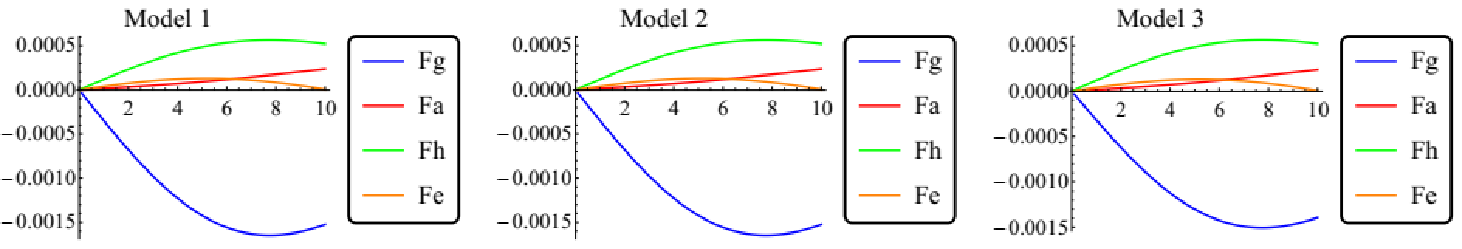,width=1\linewidth}
\caption{The Variation of hydrostatic force ($F_h$), gravitational force ($F_g$), anisotropic force $(F_a)$ and electric force $F_e$ under consideration of different viable $f(\mathcal{G})$ models.}\label{eqb}
\end{figure}
By using these definitions with the values of different parameters from Table \ref{table:1}, we check the variations of these forces and their hydrostatic equilibrium, as shown in Fig. \ref{eqb}.\\
The left plot shows the evolution of these forces in background of  first model, the middle one is for second model and the right plot describe the variation of these forces because of third model.
It is clear from Fig. \ref{eqb}, that the electric force has a very negligible effect in this balancing mechanism.
\subsection{Stability Analysis}
\begin{figure} \centering
\epsfig{file=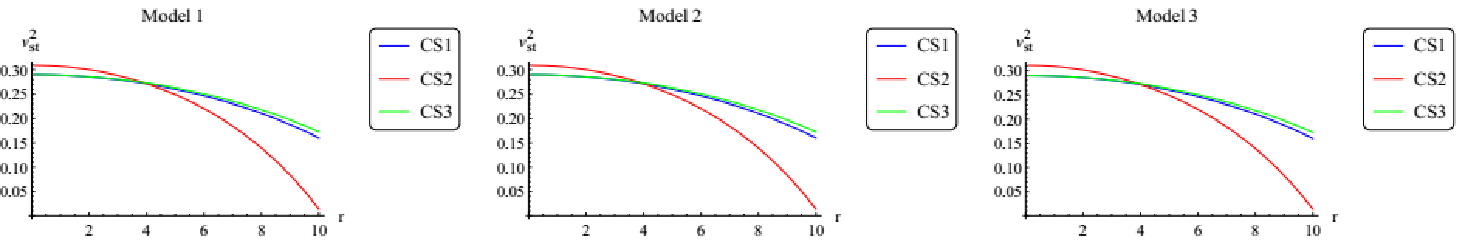,width=1\linewidth}
\caption{Variations of $v_{st}^2$ for different viable $f(\mathcal{G})$ gravity models.}\label{vst}
\end{figure}

In this section, we investigate the stability of the interior of stars under modified $f(\mathcal{G})$ theory. For the
mathematical modeling of compact stellar structures, it is to be noted that only those stellar models are significant which are stable against the variations. Hence, the role of stability is very crucial and burning issue in the modeling of compact objects. The stability of stellar structure has been studied by many researches. Here we adopt the techniques which is based on the concept of overturning (or cracking) \cite{ref30}. According to this, the radial speed of sound $v_{sr}^{2}$ as well as transverse speed of sound $v_{st}^{2}$ must be in the range of a closed interval [0, 1] to preserve the causality condition and for stability the necessary condition $0\le v_{sr}^{2}-v_{sr}^{2}\le1$ should be obeyed.\\
The the radial and transverse speeds is defined as
$$\frac{{d{p_r}}}{{d\rho }} = v_{sr}^2$$
and
$$\frac{{d{p_t}}}{{d\rho }} = v_{st}^2$$
In our case, $v_{sr}^2\sim 1/3$ and $v_{sr}^2$ is plotted in fig. \ref{vst}, which obey the condition $0 \le v_{sr}^2 \le 1$ and $0 \le v_{st}^2 \le 1$ which is the indication of causality preservation within these charged compact stars.
\begin{figure} \centering
\end{figure}
\begin{figure} \centering
\epsfig{file=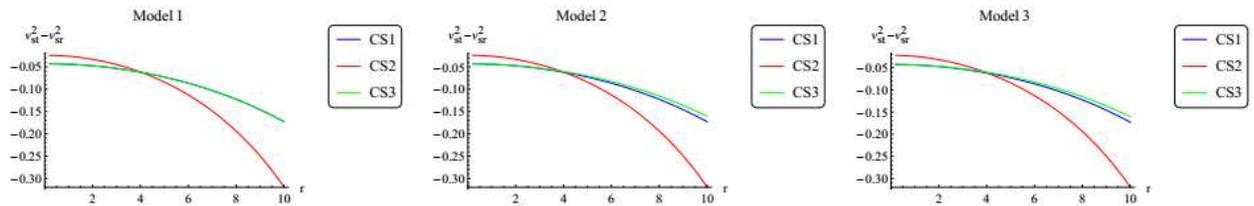,width=1\linewidth}
\caption{Variations of $v_{st}^2 - v_{sr}^2$ for different viable $f(\mathcal{G})$ gravity models.}\label{vstmvsr}
\end{figure}
Similarly, for stability, we plot $v_{st}^2-v_{sr}^2$ as shown in Fig. \ref{vstmvsr}. It is to be noted that all of our charged stellar structures under consideration of different viable $f (\mathcal{G})$ models obey the constraint:
$$0< |v_{st}^2-v_{sr}^2|<1$$
We concluded that the stability is attained in $f(\mathcal{G})$ gravity models for three considered strange candidate stars, $Vela X - 1$, $SAX J 1808.43658$, and $4U 1820-30$.
\subsection{EoS Parameter}
\begin{figure} \centering
\epsfig{file=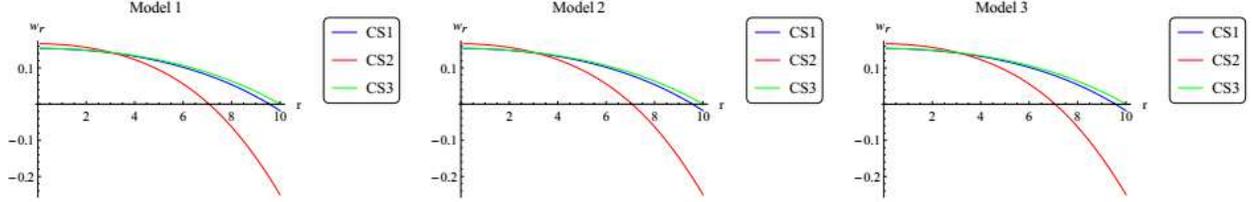,width=1\linewidth}
\caption{Variations of radial EoS parameter for different viable $f(\mathcal{G})$ gravity models.}\label{wr}
\end{figure}

\begin{figure} \centering
\epsfig{file=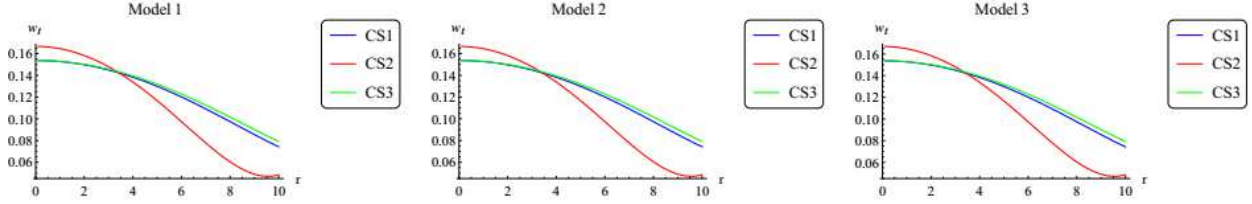,width=1\linewidth}
\caption{Variations of the transverse EoS parameter for different viable $f(\mathcal{G})$ gravity models.}\label{wt}
\end{figure}

Now for anisotropic stresses, there are two equation of state parameters, written as
$$w_r= \frac{p_r}{\rho}$$
and
$$w_t =\frac{p_t}{\rho}$$
For a radiation dominant era, equation of state parameters must lie between 0 and 1. More precisely, $0<w_r<1$ and $0<w_t<1$. Here, we check the evolution of EoS parameters for three different charged stars and their behavior are shown graphically in Fig. \ref{wr} and \ref{wt}.\\
We can see that both $w_r$ and $w_t$ lies in given range.
\subsection{ Mass Radius Relationship, Compactness and Redshift Analysis}
\begin{figure} \centering
\epsfig{file=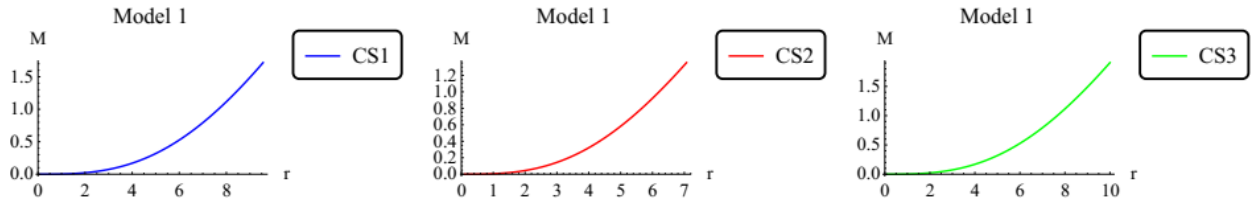,width=1\linewidth}
\caption{Variations of the mass function for different charged compact stars.}\label{mass}
\end{figure}

\begin{figure} \centering
\epsfig{file=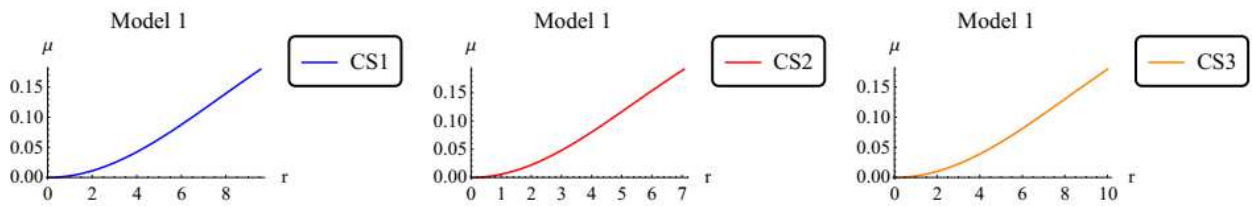,width=1\linewidth}
\caption{Variations of the compactness for different charged compact stars.}\label{compact}
\end{figure}

The mass of charged compact stars can be written as
\begin{equation}
m(r) = \int\limits_0^r {4\pi {{r'}^2}\rho dr'}
\end{equation}
Here we know that mass $m$ is function of $r$ and $m(r=0)=0$ while $m(r=R)=M$. The variation in masses of charged compact stars are shown in Fig. \ref{mass}.
We see that the mass is regular at core because it is directly proportional to radial distance e.g.  $m(r)\rightarrow0$ for $r\rightarrow0$. The maximum mass is attained at $r=R$, as shown in fig. \ref{mass}. The mass radius relation is also compatible with the study of neutron stars under $f(G)$ gravity \cite{ref50a}.\\
Furthermore, the compactness, $\mu$ can be define as
\begin{equation}
\mu (r) = \frac{1}{r}\int\limits_0^r {4\pi {{r'}^2}\rho dr'}
\end{equation}
The compactness for three different strange stars are shown in Fig. \ref{compact}. similarly, the Redshift, $Z_s$ can be define for compact object
$$Z_s = {\left( {1 - 2\mu } \right)^{ - \frac{1}{2}}} - 1$$
The bound over $Z_s\leq2$.
\begin{figure} \centering
\epsfig{file=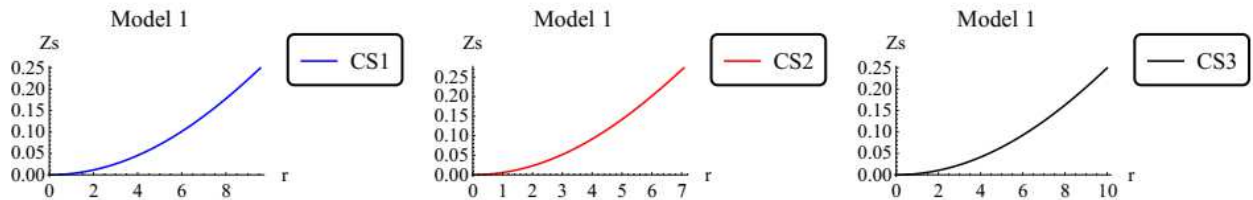,width=1\linewidth}
\caption{Variations of the Redshift for different charged compact stars.}\label{zs}
\end{figure}
In our case, wee check the variation in redshift from the core to surface of stars. These evolution are shown with the help of plots, as given in Fig. \ref{zs}.
\subsection{The Measurement of Anisotropy}
\begin{figure} \centering
\epsfig{file=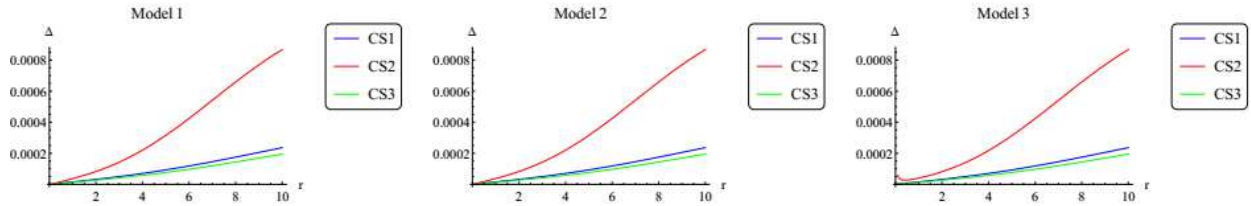,width=1\linewidth}
\caption{Variations of anisotropic measure $\Delta$ with respect to the radial.}\label{ani}
\end{figure}

In modeling of relativistic stellar interior structures, it is important to discuss the anisotropicity or anisotropy which  is defined as
\begin{equation}
\Delta  = \frac{2}{r}({p_t} - {p_r})
\end{equation}
We check the anisotropy for three different charged strange stars under consideration of three viable models in $f(\mathcal{G})$ gravity. After plugging the constant values with these models, we plot the anisotropy and get that $\Delta > 0$ e.g. $p_t > p_r$. This implies that the anisotropy is directed outward for all three stars. These plots are shown in Fig. \ref{ani}. It is important to note that $\Delta\rightarrow 0$ at $r \rightarrow 0$ and becomes monotonically
increasing outwards with the increase of $r$ near the surface of the star.
\subsection{Electric field and Charge}
\begin{figure} \centering
\epsfig{file=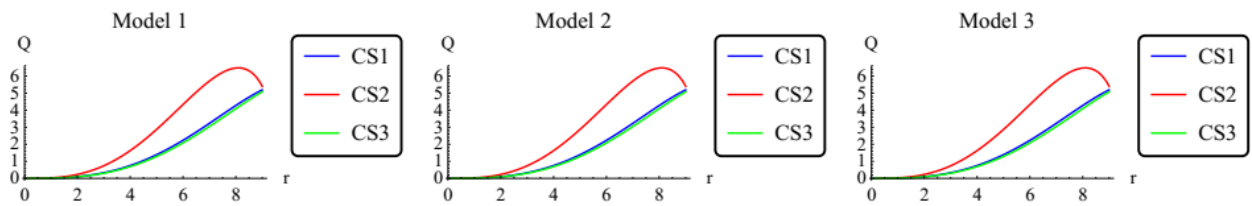,width=1\linewidth}
\caption{Variations of electric charge $Q$ with respect to the radial.}\label{charg}
\end{figure}

\begin{figure} \centering
\epsfig{file=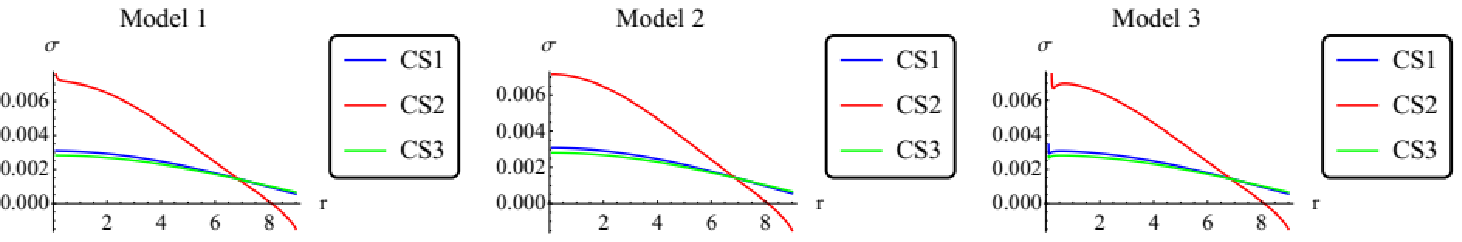,width=1\linewidth}
\caption{Variations of surface charge density $\sigma$ with respect to the radial.}\label{scharge}
\end{figure}

\begin{figure} \centering
\epsfig{file=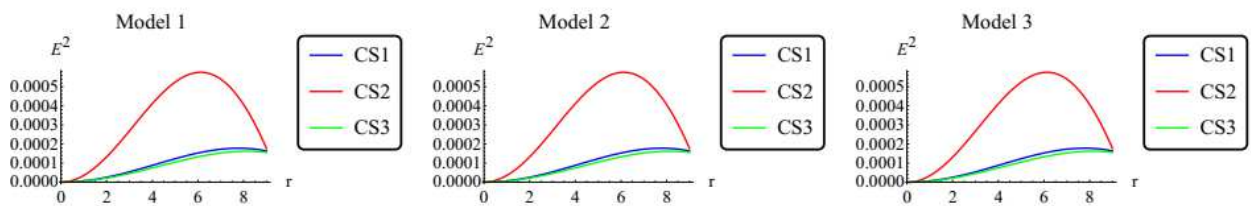,width=1\linewidth}
\caption{Variations of electric field square $E^2$ with respect to the radial.}\label{electric}
\end{figure}

We observed that the electric charge on the boundary for star 1 is $6.459234\times10^{20}C$, for star 2 $6.7510838\times10^{20}C$ and for star 3 $6.7460087\times10^{20}C$ and zero at the core of these stars under consideration of model 1. The charge profile is monotonically increasing away from the center, as shown in Fig. \ref{charg}.\\
Furthermore, the electric charge density is monotonically decreasing outward and is maximum at the center of these stars as shown in Fig. \ref{scharge}.\\
Similarly, the behavior of electric field intensity $E^2$ is also discuss and their variation for different stars are shown in Fig. \ref{electric}.\\
From these plots, we conclude that the core of these stars contain $Q(r=0)=0$, $\sigma(r=0)=\sigma_0$ and $E^2(r=0)=0$ while at surface of these stars $Q(r=R)=Q_m$ $\sigma(r=R)=0$ and $E^2(r=R)=E^2_m$ where $\sigma_0$, $Q_m$  and $E^2_m$is the maximum charge density, charge and electric field intensity.\\
The stellar structure formation in the background of modified gravity are comparatively higher contraction in the collapsing rate of spherical systems at its initial stages unlike GR.
The extra curvature (non-gravitational fluid) on the existence of compact structure could lead arena of having relatively more compact stars than in GR.\\ Similarly, the influences of these additional dark source terms on mass radius relationships for compact stars predict more massive relativistic systems with comparatively smaller radii than in GR. Perhaps, the calculated apparent masses of neutron star models in modified gravity are more massive star with smaller radii than in GR. Such type of investigations could provide theoretical well-consistent way to handle and study classes of massive and super massive structures at large scales.
\section{Summary}
\vspace{0.5cm}
It has been attracting challenge to find the correct model for charged realistic geometry of interior compact objects not only in general relativity but also in extended theories of gravity like $f(\mathcal{G})$ gravity. For this purpose, we have considered the three-different observed compact stars, labeled as Vela X - 1, SAX J 1808.4-3658, and 4U 1820-30. Our desire is to study the real composition of these compact objects in their central regions under consideration of three different viable models.\\
We have investigated several aspects of compact stars in the regime of $f(\mathcal{G})$ gravity with the anisotropic matter content under Einstein Maxwell spacetime. We have utilized the solutions for the metric function suggested by Krori-Barua for a spherical compact object whose arbitrary constants are calculated across the boundary of interior and exterior geometry. The values of these arbitrary constants are determined with the help of charge, mass, and radius of any compact object. We have used three different strange candidate stars with their experimental observational data to study the effects of additional degree of freedom coming from modified gravity theories. For this purpose, we used three different viable models in $f(\mathcal{G})$ gravity. By using these models along with calculated values for three different stars, we have plotted the relevant quantities like variation of anisotropic stress and energy density against radial distance. It is found that the energy density is very high at core of these stars and gradually decreases with the increasing radius, thereby indicating the high compactness structures of these stellar interiors.\\
We concluded our discussion as:
\begin{itemize}
  \item The variation in energy density and both radial as well as transverse stress are positive throughout these charged stars configurations.
  \item The radial derivative of density and anisotropic pressures ($d\rho/dr$, $dP_r/dr$, $dP_t/dr$) remains negative and for $r=0$, these values vanish which confirm the density and anisotropic stress maximum value at core.
  \item All the energy conditions are well satisfied which show the realistic matter content.
  \item Both the radial and transverse sound speed remains within the bounds, which mean the causality condition is obeyed.
  \item All these stars are stable.
  \item Both radial and transverse EoS parameters lie in the range of 0 and 1.
  \item The isotropy remains positive throughout these charged stars.
  \item The distribution of charges increase from central to surface of stars.
  \item Electric field intensity is maximum at the surface of these stars.
\end{itemize}
We see that $f(\mathcal{G})$ gravity is much attractive in the study of compact stars. In this sense, to our knowledge, the existence and study of different charged stars and particle physics inside their highly dense cores compelled the researchers for more genuine solutions of field equations. Similarly, the study of compact charged stellar configuration and better observational data on the mass radius relation have the potential to exclude a larger region of the parameter space of alternative theories. There is of course the possibility that theoretical and observational work may give us a direction on how to modify general relativity to make it compatible with the standard model of particle physics, which would be even more exciting.


\begin{thebibliography}{72}
\bibitem{refa1}
K.~Koyama, Reports on Progress in Physics \textbf{79}(4), 046902 (2016)
\bibitem{refa2}
A.~de~la Cruz-Dombriz, D.~S{\'a}ez-G{\'o}mez, Entropy \textbf{14}(9), 1717
  (2012)
\bibitem{refa3}
K.~Bamba, S.~Nojiri, S.D. Odintsov, arXiv preprint arXiv:1302.4831  (2013)
\bibitem{refa4}
K.~Bamba, S.D. Odintsov, arXiv preprint arXiv:1402.7114  (2014)
\bibitem{refa4a}
K. Bamba, S.D. Odintsov, Symmetry 7, 220 (2015). 502
arXiv:1503.00442 [hep-th]
\bibitem{refa5}
Z.~Yousaf, K.~Bamba, et~al., Physical Review D \textbf{93}(6), 064059 (2016)
\bibitem{refa6}
Z.~Yousaf, K.~Bamba, M.Z.u.H. Bhatti, Physical Review D \textbf{93}(12), 124048
  (2016)
\bibitem{refa7}
T.P. Sotiriou, V.~Faraoni, Reviews of Modern Physics \textbf{82}(1), 451 (2010)
\bibitem{refa7aaa}
S.~Nojiri, S.D. Odintsov, International Journal of Geometric Methods in Modern
  Physics \textbf{4}(01), 115 (2007)
\bibitem{ref1}
K.~Bamba, S.~Capozziello, S.~Nojiri, S.D. Odintsov, Astrophysics and Space
  Science \textbf{342}(1), 155 (2012)
\bibitem{ref2}
S.~Capozziello, V.~Faraoni, \emph{Beyond Einstein gravity: A Survey of
  gravitational theories for cosmology and astrophysics}, vol. 170 (Springer
  Science \& Business Media, 2010)
\bibitem{ref3}
G.~Dvali, G.~Gabadadze, M.~Porrati, Physics Letters B \textbf{485}(1-3), 208
  (2000)
\bibitem{ref4}
H.A. Buchdahl, Monthly Notices of the Royal Astronomical Society
  \textbf{150}(1), 1 (1970)
\bibitem{ref5}
S.~Capozziello, M.~De~Laurentis, S.~Odintsov, A.~Stabile, Physical Review D
  \textbf{83}(6), 064004 (2011)
\bibitem{ref6}
S.~Nojiri, Phys. Rep. \textbf{505}, 59 (2011)
\bibitem{ref7}
T.~Harko, Monthly Notices of the Royal Astronomical Society \textbf{413}(4),
  3095 (2011)
\bibitem{ref8}
D.~Lovelock, Journal of Mathematical Physics \textbf{12}(3), 498 (1971)
\bibitem{ref10}
M.~Sharif, A.~Ikram, Journal of Experimental and Theoretical Physics
  \textbf{123}(1), 40 (2016)
\bibitem{ref11}
M.F. Shamir, M.~Ahmad, The European Physical Journal C \textbf{77}(1), 55
  (2017)
\bibitem{ref12}
M.~Shamir, Mod. Phys. Lett. A \textbf{32}, 1750086 (2017)
\bibitem{ref14}
G.~Cognola, Phys. Rev. D \textbf{73}, 084007 (2006)
\bibitem{ref15}
E.~Elizalde, R.~Myrzakulov, V.V. Obukhov, D.~S{\'a}ez-G{\'o}mez, Classical and
  Quantum Gravity \textbf{27}(9), 095007 (2010)
\bibitem{ref13}
R.~Kitano, Y.~Nomura, Physics Letters B \textbf{631}(1-2), 58 (2005)
\bibitem{ref16}
A.~De~Felice, T.~Tanaka, Progress of Theoretical Physics \textbf{124}(3), 503
  (2010)
\bibitem{ref17}
A.~De~Felice, T.~Suyama, Journal of Cosmology and Astroparticle Physics
  \textbf{2009}(06), 034 (2009)
\bibitem{ref47}
M.~Visser, Science \textbf{276}(5309), 88 (1997)
\bibitem{ref48}
M.~Visser, Physical Review D \textbf{56}(12), 7578 (1997)
\bibitem{ref49}
J.~Santos, J.~Alcaniz, M.J. Rebou{\c{c}}as, Physical Review D \textbf{74}(6),
  067301 (2006)
\bibitem{ref50}
J.~Santos, J.~Alcaniz, N.~Pires, M.J. Reboucas, Physical Review D
  \textbf{75}(8), 083523 (2007)
\bibitem{ref52}
J.~Santos, J.~Alcaniz, M.~Rebou{\c{c}}as, N.~Pires, Physical Review D
  \textbf{76}(4), 043519 (2007)
\bibitem{ref51}
A.~Sen, R.J. Scherrer, Physics Letters B \textbf{659}(3), 457 (2008)
\bibitem{ref53}
Y.~Gong, A.~Wang, Q.~Wu, Y.Z. Zhang, Journal of Cosmology and Astroparticle
  Physics \textbf{2007}(08), 018 (2007)
\bibitem{ref54}
Y.~Gong, A.~Wang, Physics Letters B \textbf{652}(2-3), 63 (2007)
\bibitem{ref55}
J.~Santos, J.~Alcaniz, M.~Reboucas, F.~Carvalho, Physical Review D
  \textbf{76}(8), 083513 (2007)
\bibitem{ref56}
J.~SANTOS, M.J. REBOU{\c{C}}AS, J.S. ALCANIZ, International Journal of Modern
  Physics D \textbf{19}(08n10), 1315 (2010)
\bibitem{ref18}
F.W. Hehl, C.~Kiefer, R.J. Metzler, \emph{Black Holes: Theory and Observation:
  Proceedings of the 179th WE Heraeus Seminar Held at Bad Honnef, Germany,
  18--22 August 1997}, vol. 514 (Springer Science \& Business Media, 1998)
\bibitem{ref19}
Z.~Yousaf, M.~Sharif, M.~Ilyas, M.~Bhatti, The European Physical Journal C
  \textbf{77}(10), 691 (2017)
\bibitem{ref19a}
M.Z.u.H. Bhatti, M.~Sharif, Z.~Yousaf, M.~Ilyas, International Journal of
  Modern Physics D \textbf{27}(04), 1850044 (2018)
\bibitem{ref19b}
M.~Ilyas, Z.~Yousaf, M.~Bhatti, B.~Masud, Astrophysics and Space Science
  \textbf{362}(12), 237 (2017),\\
Z. Yousaf, M.Z. Bhatti, M. Ilyas, Eur. Phys. J. C \textbf{78}, 307 (2018)
\bibitem{ref21}
R.L. Bowers, E.~Liang, The Astrophysical Journal \textbf{188}, 657 (1974)
\bibitem{ref25}
K.N. Singh, N.~Pant, Govender, M, Chinese Physics C \textbf{41}(1), 015103
  (2017)
\bibitem{ref26}
P.~Bhar, Astrophysics and Space Science \textbf{356}(2), 309 (2015)
\bibitem{ref27}
P.~Bhar, The European Physical Journal C \textbf{75}(3), 123 (2015)
\bibitem{ref28}
K.N. Singh, N.~Pant, Astrophysics and Space Science \textbf{358}(2), 44 (2015)
\bibitem{ref29}
K.~Krori, J.~Barua, Journal of Physics A: Mathematical and General
  \textbf{8}(4), 508 (1975)
\bibitem{ref43}
H.J. Schmidt, Physical Review D \textbf{83}(8), 083513 (2011)
\bibitem{ref44}
K.~Bamba, S.D. Odintsov, L.~Sebastiani, S.~Zerbini, The European Physical
  Journal C \textbf{67}(1-2), 295 (2010)
\bibitem{ref30}
L.~Herrera, Physics Letters A \textbf{165}(3), 206 (1992)
\bibitem{ref50a}
A.V. Astashenok, S.~Capozziello, S.D. Odintsov, Journal of Cosmology and
  Astroparticle Physics \textbf{2015}(01), 001 (2015)
\end{thebibliography}
\end{document}